\documentclass[12pt]{article}
\usepackage{amsmath, amsfonts, amsthm, amssymb,epsfig}
\usepackage[english]{babel}
\usepackage{authblk}
\title{Constraining exotic interactions}

\author[1]{Filip Ficek\footnote{Corresponding author E-mail:~\textsf{filip.ficek@uj.edu.pl}}}
\author[2,3,4]{Dmitry Budker}
\affil[1]{Institute of Physics, Jagiellonian University, \L{}ojasiewicza 11, 30-348 Krak\'{o}w, Poland}
\affil[2]{Helmholtz Institute Mainz, Johannes Gutenberg University, 55099 Mainz, Germany}
\affil[3]{Department of Physics, University of California at Berkeley, Berkeley, California 94720-7300, USA}
\affil[4]{Nuclear Science Division, Lawrence Berkeley National Laboratory, Berkeley, California 94720, USA}

\begin{document}
\maketitle

\begin{abstract}
Beyond-the-standard-model interactions mediated by an exchange of virtual ``new'' bosons result in a finite set of possible effective interaction potentials between standard-model particles such as electrons and nucleons. We discuss the classification of such potentials and briefly review recent experiments searching for such exotic interactions at spatial scales from sub-nanometers to tens of thousand kilometers.
\end{abstract}

\section{Introduction}\label{sec:intro}
Modern physics acknowledges the existence of four fundamental interactions -- strong, weak, electromagnetic, and gravitational. They vary in strengths and ranges, and for different physical systems some of them may be more important than the others (for example strong interactions inside baryons or gravitational interactions in the galaxy). In spite of the fact that there is no direct proof of existence of any other fundamental interaction (although there are many observations suggesting their existence, as we discussed in the next section), there is in principle no argument ruling out such a possibility. Instead, one can only constrain strengths of such hypothetical interactions using precise experimental measurements. 

In this short article, which is intended as a brief introduction rather than a comprehensive review (see Ref. \cite{Saf17} for a review on results of searches for exotic interactions based on the techniques of atomic, molecular, and optical physics), we present the basic ideas underlying searches for hypothetical interactions called ``exotic interactions'', as they may be present in extensions of the Standard Model. We begin with a presentation of a non-relativistic framework used to deal with fundamental interactions carried by spin-0 and spin-1 bosons at low-energy scales and then we explore some of the systems used to give such constraints at various scales.

\section{Exotic potentials}\label{sec:ex}
At this moment we know that every fundamental interaction, except gravity for which a satisfactory quantum theory is not yet known, is carried by some interacting bosons -- photons for electromagnetic, gluons for strong, and $Z^0/W^\pm$ bosons for weak interactions. We suspect that exotic interactions also would be carried by some, yet undiscovered bosons. Many modern physics puzzles, such as the nature of dark matter \cite{Ber05} and dark energy \cite{Fri03,Fla09}, the strong-CP problem \cite{Moo84}, or the hierarchy problem \cite{Gra15}, may be explained by Beyond Standard Model theories predicting existence of such new bosons. The Examples include axions \cite{Wei78,Wil78,Din81,Shi80,Kim79,Zhi80}, familons \cite{Wil82,Gel83}, majorons \cite{Gel81,Chi81}, new spin-0 or spin-1 gravitons \cite{Sch79,Nev80,Nev82,Car94}, Kaluza-Klein zero modes in string theory \cite{Svr06}, paraphotons \cite{Oku82,Hol86,Dob05}, and new $Z'$ bosons \cite{Bou83,App03,Dzu17}. Despite the different nature of all these particles and the reasons the corresponding models were proposed, interactions they carry may be described within one, general framework introduced by Moody and Wilczek \cite{Moo84} and expanded by Dobrescu and Mocioiu \cite{Dob06}. We follow the lines of Ref. \cite{Dob06} in order to introduce this framework.

\begin{figure}
  \caption{A Feynman diagram of two interacting fermions.}
  \centering
\includegraphics[width=0.4\textwidth]{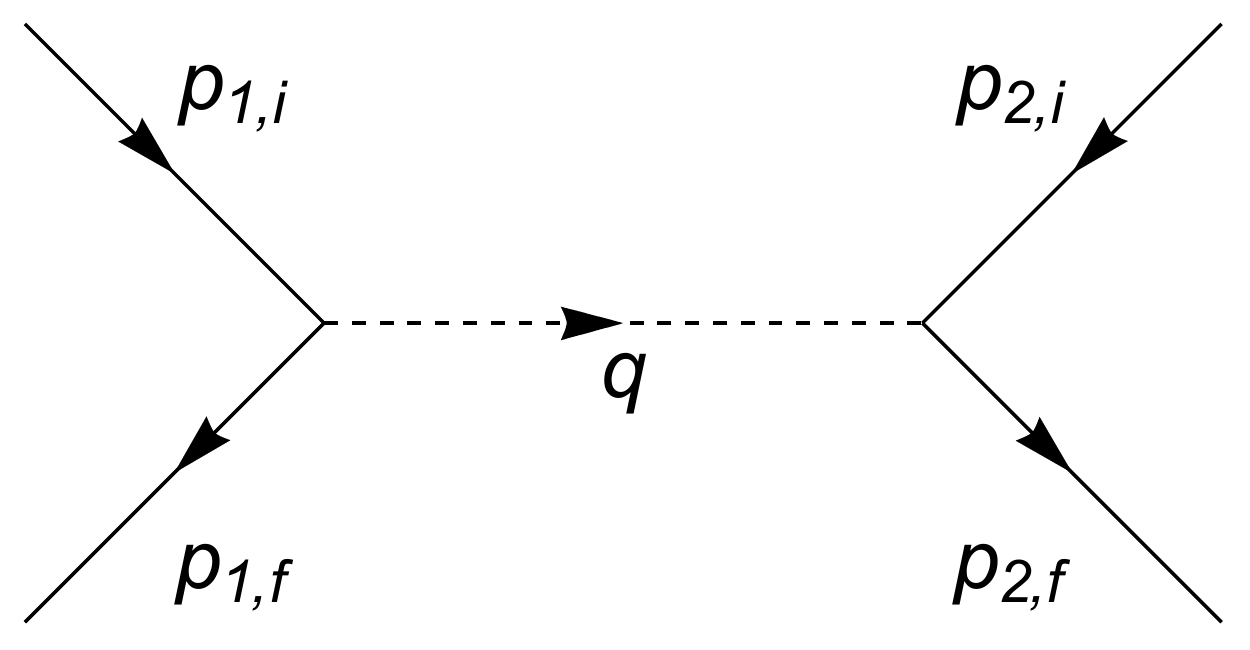}\label{fig:graph}
\end{figure}

Let us consider an interaction between two fermion particles mediated by a light boson with mass $m_0$, as shown in Fig.\ \ref{fig:graph}. The particle 1 with initial momentum $p_{1,i}$ interacts with the particle 2 having initial momentum $p_{2,i}$. The interaction is carried by a boson with momentum $q$ and as a result, the two particles carry out momenta $p_{1,f}$ and $p_{2,f}$, respectively. Let us consider this event in a center of mass of this system. Then due to the energy-momentum conservation, all the information about the kinematics of the collision is contained in the two momenta, $p_{1,i}$ and $p_{1,f}$. We are interested in low-energy interactions, as the higher-order relativistic corrections are negligible at atomic and larger scales which are our points of interest. Then the rest mass dominates the particle energy and we may consider just the spatial parts $\mathbf{p}_{1,i}$ and $\mathbf{p}_{1,f}$ of the momenta $p_{1,f}$ and $p_{2,f}$. We construct out of them the mean momentum of one of the particles $\mathbf{P}$ and the difference in initial and final momenta for this particle $\mathbf{q}$ (which are equal in magnitudes to the respective quantities for the second particle):
\begin{align} 
\mathbf{P}&=\frac{1}{2}(\mathbf{p}_{1,i}+\mathbf{p}_{1,f}),\\
\mathbf{q}&=\mathbf{p}_{1,i}-\mathbf{p}_{1,f}.
\label{eqn:fifth}
\end{align}
If spins of the particles are $\mathbf{s}_1$ and $\mathbf{s}_2$, respectively, then all the information about the collision is carried by four vectors: $\mathbf{P}$, $\mathbf{q}$, $\mathbf{s}_1$, and $\mathbf{s}_2$. Dobrescu and Mocioiu showed \cite{Dob06}, that any scalar constructed from these vectors can be presented as a linear combination of only sixteen base scalars $\mathcal{O}_i$ with coefficients depending on $\mathbf{P}^2$ and $\mathbf{q}^2$. We consider them as interaction potentials written in momentum space. One of them does not include any of spins, i.e. is algebrically equivalent to $1$. Interactions described by this potential are usually called fifth forces \cite{Fis87} and are often considered in a context of modifications of Newtonian gravity \cite{Fis99}. Interactions coming from the other fifteen potentials are spin-dependent and they divide into two groups: ones that do not include $\mathbf{P}$, called velocity-independent or static, and ones that include $\mathbf{P}$, called velocity-dependent.

Potentials in momentum space can be easily converted to a position space. As an example we may consider a potential labeled in Ref. \cite{Dob06} as $\mathcal{O}_3$. In momentum space we may write it as
\begin{align} 
\mathcal{O}_3=\frac{1}{m_e^2}(\mathbf{s}_1\cdot\mathbf{q})(\mathbf{s}_2\cdot\mathbf{q}),
\label{eqn:o3}
\end{align}
where the factor containing the electron mass $m_e$ is introduced for dimensional reasons. It may be rewritten into position space by performing a Fourier transform with an appropriate propagator $\mathcal{P}$:
\begin{align} 
\mathcal{V}_i=-\int \frac{d^3 q}{(2\pi)^3} e^{i\mathbf{q}\cdot\mathbf{r}}\mathcal{P}\mathcal{O}_i.
\label{eqn:ov}
\end{align}
We are considering the exchange presented in Fig.\ \ref{fig:graph} within a Lorentz invariant quantum field theory, which fixes the form of the propagator to $\mathcal{P}=-\frac{1}{\mathbf{q}^2+m_0^2}$ \cite{Dob06}. In principle other forms are possible, for example, coming from the exchange of two bosons instead of one or from Lorentz-symmetry violation. In the Lorentz-invariant, single-boson-exchange framework, we get an exotic potential of the form 
\begin{align} 
\mathcal{V}_3=-\frac{1}{4 \pi} &\left[\mathbf{s}_1\cdot\mathbf{s}_2\left(\frac{1}{\lambda r^2}+\frac{1}{r^3}+\frac{4\pi}{3}\delta^3(r) \right)\right.\nonumber\\
&\left.-(\mathbf{s}_1\cdot\mathbf{r})(\mathbf{s}_2\cdot\mathbf{r})\left(\frac{1}{\lambda^2 r^3}+\frac{3}{\lambda r^4}+\frac{3}{r^5}\right)\right]e^{-r/\lambda},
\label{eqn:v3a}
\end{align}
where $\mathbf{r}$ is a vector connecting the interacting particles, $r$ is its length (distance between the particles), and $\lambda=\hbar/m_0 c$ is the Compton length of the interaction-mediating boson. We have included here a sign correction recently introduced by Daido and Takahashi \cite{Dai17}.

To obtain a final form of the exotic potential in position space we need to give it a proper dimension by inserting an overall constant. Dimensional analysis yields $\hbar^3/m_e^2 c$, where $\hbar$ is the reduced Planck constant and $c$ is the speed of light, as a correct combination. Additionally we put a dimensionless coupling constant $f^{12}_3$ which represents the strength of this interaction and may depend on interacting particles (hence the index 12 referring to particle 1 and particle 2). The coupling coefficients $f^{12}_3$ (often written as $g^1_3 g^2_3/4\pi\hbar c$) are determined by experimental searches. In the end we get
\begin{align} 
V_3=-f^{12}_3 \frac{\hbar^3}{4 \pi m_e^2 c} &\left[\mathbf{s}_1\cdot\mathbf{s}_2\left(\frac{1}{\lambda r^2}+\frac{1}{r^3}+\frac{4\pi}{3}\delta^3(r) \right)\right.\nonumber\\
&\left.-(\mathbf{s}_1\cdot\mathbf{r})(\mathbf{s}_2\cdot\mathbf{r})\left(\frac{1}{\lambda^2 r^3}+\frac{3}{\lambda r^4}+\frac{3}{r^5}\right)\right]e^{-r/\lambda}.
\label{eqn:v3b}
\end{align}
This potential, usually called a pseudovector dipole-dipole potential, was for the first time considered by Moody and Wilczek in Ref. \cite{Moo84} and may be associated for example with an exchange of an axion. Another often consider dipole-dipole potential comes from an exchange of an axial-vector particle and has the form of
\begin{align} 
V_2=f^{12}_2 \frac{\hbar c}{\pi} (\mathbf{s}_1\cdot\mathbf{s}_2)\frac{e^{-r/\lambda}}{r}.
\label{eqn:v2}
\end{align}

The procedure that gave us Eq.\ (\ref{eqn:v3a}) may be repeated for the remaining fifteen scalars, although one should be cautious when dealing with velocity-dependent potentials. After performing the Fourier transformation for these operators, the authors of Ref. \cite{Dob06} kept vectors $\mathbf{P}$ as variables instead of changing them into operators related to gradient. As realized by M. G. Kozlov \cite{Fic17}, this gives the potentials in some kind of mixed representation, which may be used at the laboratory scale (as considered in Ref. \cite{Dob06}), but is not suitable for atomic scales. As an example, one may take the momentum space operator
\begin{align} 
\mathcal{O}_8=\frac{1}{m_e^2}(\mathbf{s}_1\cdot\mathbf{P})(\mathbf{s}_2\cdot\mathbf{P})
\label{eqn:o8}
\end{align}
and perform Fourier transformation obtaining \cite{Dob06}
\begin{align} 
\mathcal{V}_8=\frac{1}{4 \pi r} (\mathbf{s}_1\cdot\mathbf{P})(\mathbf{s}_2\cdot\mathbf{P}) e^{-r/\lambda}.
\label{eqn:v8a}
\end{align}
In order to get correct position space forms of velocity-dependent potentials, one needs to perform an additional antisymmetrization, as described in Ref. \cite{Fic17}. Then Eq.\ (\ref{eqn:v8a}) transforms to
\begin{align} 
V_8=&-f^{12}_8 \frac{\hbar^3}{4\pi m_e^2 c}\left\{\mathbf{s}_1\cdot\left(\frac{m_1}{m_1+m_2}\nabla_2-\frac{m_2}{m_1+m_2}\nabla_1\right),\right.\nonumber\\
&\left.\left\{\mathbf{s}_2\cdot\left(\frac{m_1}{m_1+m_2}\nabla_2-\frac{m_2}{m_1+m_2}\nabla_1\right),{e^{-r/\lambda}}{r}\right\}\right\},
\label{eqn:v8b}
\end{align}
where $\{\cdot,\cdot\}$ denotes anticommutator. The full list of potentials can be found in Ref.\ \cite{Saf17}


Let us point out that potentials obtained by the described method come from very general principles, so it is worth considering whether all of them have some physical interpretation. Recently Fadeev et.al. \cite{FadIP} performed an alternative construction of exotic potentials. One may start from the most general Lorentz-invariant Lagrangian describing interactions between standard-model fermions and spin-0 or spin-1 bosons. For example in a scalar sector such Lagrangian has the form
\begin{align} 
\mathcal{L}_\phi=\phi \bar{\psi} \left( g^s_\psi+i \gamma_5 g^p_\psi\right)\psi,
\end{align}
where $\psi$ is a fermionic field, $\phi$ is a scalar field, $\gamma_5$ is a Dirac matrix, and $g^s_\psi$, $g^p_\psi$ parameterise interaction strengths (the first one applies to $P$-even, hence $s$ as scalar, and the second to $P$-odd interactions, hence $p$, as pseudoscalar). Similar terms may be written for massless and massive spin-1 particles giving six parameters in total. Investigating the vertices of the Feynman diagram presented in Fig.\ \ref{fig:graph} with this general lagrangian yields potentials that can be considered in a nonrelativistic limit. These limits happen to be linear combinations of the potentials obtained by Dobrescu and Mocioiu, although not all sixteen of them are present. This suggests, that some of the $\mathcal{O}_i$ scalars have no physical significance. The additional result coming from this alternative approach is the fact that not all of the dimensionless coupling constants $f^{\psi\psi}_i$ are independent -- they can be expressed as combinations of $g^s_\psi$, $g^p_\psi$ and the remaining four parameters mentioned above.

Every exotic potential presented in Ref. \cite{Dob06} contains exponential factor $\exp(-r/\lambda)$ suppressing the interaction at scales higher then the Compton wavelength $\lambda$ of the mediating boson. This means, that the boson mass $m_0$ determines the characteristic scale of interaction and, as an effect, investigating different physical systems may give us constraints on exotic interactions carried by bosons with different masses. In the next section we review some of physical systems yielding constraints at different mass scales.

\section{Methods}\label{sec:met}
In this section we present several experimental methods used to obtain constraints on exotic interactions. The common idea behind them all consists in performing an experiment and then comparing its results with standard (for example QED based) theoretical predictions in order to find any deviations or at least determine the uncertainties to which the agreement between theory and observations can be established. The difference between experimental results and theoretical predictions gives us a window where some additional exotic interactions may fit (the narrower window, the more stringent the final constraints). By calculating the influence of hypothetical exotic interactions on the results of experiments we may search for them, and either find something or obtain constraints on the appropriate coupling constants. We show how this procedure works in a particular case in the next section.

As mentioned at the end of the last section, experiments performed at different scales are affected by forces mediated by bosons with different masses. Because of this, we need to utilise experiments working at various scales to properly investigate the exotic interaction parameter space. Also some experimental setups may be sensitive to different kinds of exotic interactions, such as spin-dependent or velocity-dependent, while others are not, which highlights the need for large diversity of investigated systems.

In the following sections we discuss four experimental methods yielding constraints on spin-dependent exotic interactions at various scales, from nanometers up to thousands of kilometers (or from 1 keV down to $10^{-12}$ eV, equivalently). We present the constraints on coupling constants for axial-vector and pseudoscalar dipole-dipole interactions between electrons ($|f_2^{ee}|$ and $|f_3^{ee}|$, respectively) where possible. More detailed descriptions of the results together with limits on other potentials may be found in the cited references. Apart from the experimental techniques presented below, there is a variety of others, for example, trapped ions experiments \cite{Kot15}, molecular spectroscopy \cite{Ram79, Led13}, measurements of the spin precession of atomic gases \cite{Tul13, Chu16}. There are also new ideas such as, for example, a scanning tunneling microscopy \cite{Luo17}. A comprehensive review of searches of exotic interactions with atomic and molecular experiments can be found in Ref.\ \cite{Saf17}. Also one can find a list of constraints on some coupling constants at different scales in The Review of Particle Physics \cite{PDG18}.

Let us also mention, that the strength of exotic forces is constrained not only with earthbound experiments. Such hypothetical interactions would influence many astrophysical processes and their impact should be visible in astronomical observations. By comparing the predictions regarding such observables as red-giant cooling rate \cite{Hax91, Raf95} or strength of neutrino flux from supernovae \cite{Eng90} with observational evidence, it is possible to put strong limits on the possible new interactions. However, one has to keep in mind, that some of these results heavily rely on the astrophysical models employed.

\subsection{Atomic spectroscopy}
Precision levels achieved by modern atomic spectroscopy and QED-based theoretical calculations, together with a good agreement between them, let us obtain stringent constraints on exotic interactions at the atomic scale. The diversity of exotic atoms permits us to search for exotic interactions strengths between various particles. The existing results include limits on interactions at the atomic scale between two electrons (from helium \cite{Fic17}), electron and positon (from positronium \cite{Kot15}), electron and antimuon (from muonium \cite{Kar10, Kar11}), or electron and antiproton (from antiprotonic helium \cite{Fic18}). Details regarding searches for exotic interactions vary from system to system, so as an example in the remaining part of this section we focus on limits on exotic interactions between electrons coming from helium fine structure, as described in Ref.\ \cite{Fic17}.

Let us investigate the $n=2$ state (where $n$ is the principal quantum number) of orthohelium. It consists of a metastable state $2 ^3 S$ and a triplet of $2^3 P$ states. Transition energies between these states have been precisely measured \cite{Zhe17, Mar15}. These frequencies may be compared with QED-based calculations \cite{Pac17, Pac10} (whose precision is lower than that of experimental data) to reveal that they agree within the uncertainties. It suggests, that possible exotic interactions must fit in these uncertainties. Let us focus on one of the transitions. We want to define a quantity characterising the level of agreement between theory and experiment taking into account the uncertainties, called $\Delta E$ from now on. If we denote by $\mu$ the mean difference between its theoretical and experimental frequencies and also we define $\sigma=\sqrt{\sigma_{th}^2+\sigma^2_{exp}}$ (where $\sigma_{th}$ and $\sigma_{exp}$ are theoretical and experimental uncertainties, respectively), we may introduce $\Delta E$ as a number such that [cf. Eqs. (A1) and (A2) of Ref. \cite{Fic17} where, apart from the typo in Eq. (A2), an equivalent definition of $\Delta E$ is given]
\begin{align} 
\int_{-\Delta E}^{\Delta E} \frac{1}{\sqrt{2\pi \sigma}} e^{-(x-\mu)^2/2\sigma^2} dx=0.9.
\label{eqn:DE}
\end{align}
This number may be calculated for every transition and can be interpreted as a maximal possible energy shift caused by exotic interactions for this transition (at $90\%$ acceptance level). Let us now point out, that we can factor out the coupling constant $f_i^{ee}$ from every exotic potential $V_i$ getting $V_i=f_i^{ee} U_i$, where $U_i$ is a well defined operator. We now consider a transition between states A and B, characterised by electron wavefunctions $|\psi_A\rangle$ and $\psi_B\rangle$, respectively. Exotic potential $V_i$ shifts energy of the state A by $\langle \psi_A|V_i|\psi_A \rangle=f_i^{ee}\langle \psi_A|U_i|\psi_A \rangle$, and analogously for the state B. It means, that the total change in frequency for a transition A$\leftrightarrow$B caused by potential $V_i$ is $|\langle \psi_A|V_i|\psi_A \rangle-\langle \psi_B|V_i|\psi_B \rangle|$. This quantity cannot be larger than $\Delta E$. Connecting all these information we arrive at the final expression
\begin{align} 
|f_i^{ee}|\leq \frac{\Delta E}{|\langle \psi_A|U_i|\psi_A \rangle-\langle \psi_B|U_i|\psi_B \rangle|}.
\label{eqn:DE}
\end{align}
By performing these steps for appropriate transitions within helium fine structure, one can obtain limits $|f_2^{ee}|\leq 10^{-9}$ and $|f_3^{ee}|\leq 3\times10^{-8}$ at the scale of 1 nm (1 keV) \cite{Fic17}.

\subsection{Nitrogen-vacancy centers in diamond}
\begin{figure}
  \caption{Atomic structure of a nitrogen-vacancy center in diamond.  The gray balls are carbon atoms, the yellow ball is a nitrogen atom, while the white ball symbolises the vacancy.}
  \centering
\includegraphics[width=0.4\textwidth]{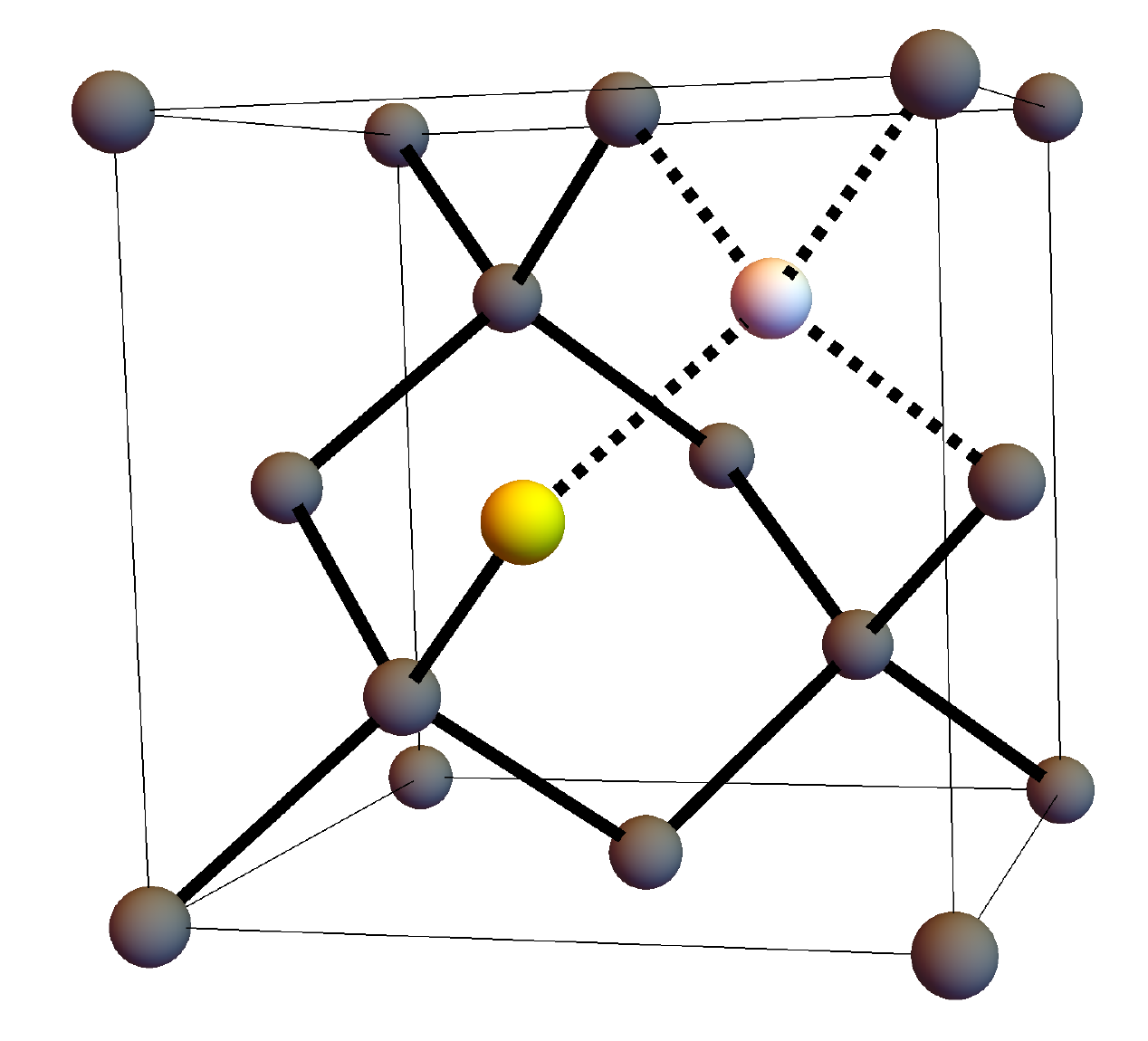}\label{fig:diamond}
\end{figure}

Nitrogen-vacancy (NV) centers are point defects in diamond structure. These occur when a pair of neighbouring carbon atoms are substituted with a single nitrogen atom and a vacancy (Fig.\ \ref{fig:diamond}). They have broad applications in quantum information, metrology, and nanotechnology \cite{Doh13}. They can also be used to measure magnetic fields with nanoscale resolution \cite{Deg17}. Because spin-dependent exotic interactions couple to the matter in a way similar to a magnetic field, this last property, together with the possibility of isolation of magnetic noise, suggests that NV centers may be used to search for exotic interactions \cite{Ron18a, Ron18b}.

In two recently published articles, Xing Rong and his collaborators described how they used a single NV centers as a quantum sensor to constrain axion-mediated monopole-dipole interactions between electron and nucleon \cite{Ron18a} and vector-mediated dipole-dipole interactions between electrons \cite{Ron18b}. Results obtained for the latter interaction allowed to obtain a limit for a dimensionless coupling constant $|f^{ee}_2|\leq 5.7 \times 10^{-19}$ at the scale of 500 $\mu$m (2.5 meV).

\subsection{Torsion balance}
\begin{figure}
  \caption{Torsion balance used in the original Cavendish experiment. Figure 1 of Ref.\ \cite{Cav98}.}
  \centering
\includegraphics[width=0.5\textwidth]{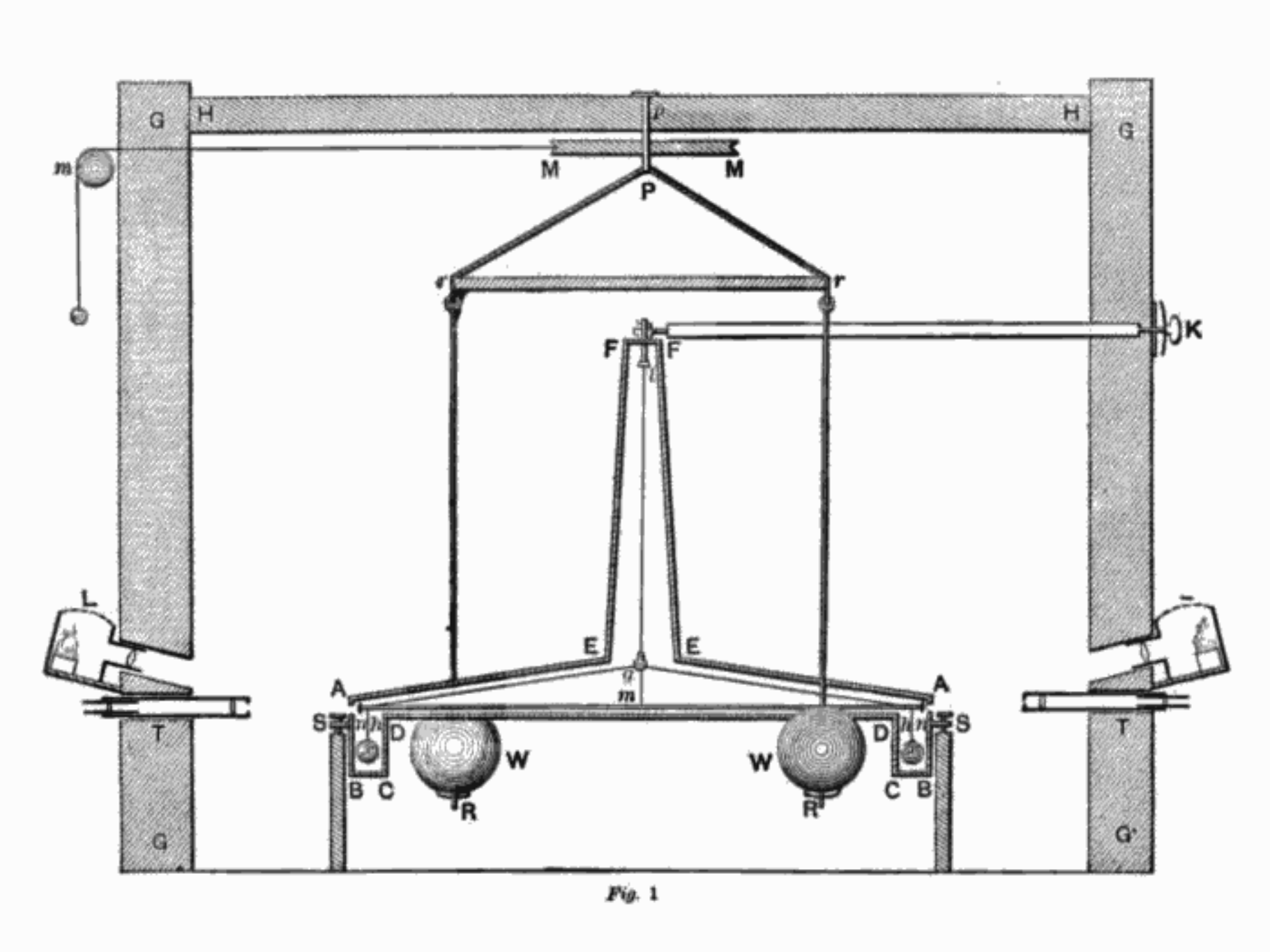}\label{fig:cavendish}
\end{figure}

In the original paper by Moody and Wilczek \cite{Moo84}, the authors proposed constraining exotic interactions with the use of techniques of experimental gravity, specifically, precise torsion-balance measurements. Such experiments are based on ideas similar to the ones behind the famous Cavendish experiment \cite{Cav98}, shown schematically in Fig.\ \ref{fig:cavendish}. Basically, two 0.73-kg lead spheres (inside $ABCD$ boxes) were attached to the opposite ends of horizontally suspended, 1.8-m wooden rod $m$ and located 23 cm away from two 158-kg lead spheres $W$ acting as weak sources of gravitational attraction. The rod with the small balls twists to the angle where a torque coming from the aforementioned gravitational force is balanced by the torque exerted by the spring. This system, initially used to find Earth's mean density \cite{Clo87} (which could be converted to the value of gravitational constant), after some changes may search for deviations from Newton's inverse--square law. Such deviations could come from yet undiscovered forces, rather then being connected to the nature of gravity. They would be results of spin-independent fifth-forces, as both source and detector in this setup are unpolarized, and such experiments may yield constraints on their strengths \cite{Fis99, Hoy04, Ade05, Upa12}.

After further modifications to the torsion balance it is also possible to search for constraints on spin-dependent exotic forces. Such setup must not only contain polarized test bodies and sources, but also should be shielded from any external magnetic fields. Examples of such apparatus come from the E{\"o}t-Wash Group at the University of Washington, where they were used to constrain various types of spin-dependent interactions, including $CP$-violating forces \cite{Hec08}, axion mediated forces \cite{Hoe11}, and spin-spin interactions between electrons at various length scales \cite{Hec13, Ter15}. The most recent results come from a system utilising a 4-cm-wide ring containing 20 magnetized segments of alternating high and low spin-density materials \cite{Ter15}. This setup allows for a great reduction of an influence of external magnetic fields, while keeping sensitivity to exotic spin-dependent forces thanks to variations in spin density. The results coming from these experiments yield constraints on the coupling constants being $|f^{ee}_2|\leq 5.1 \times 10^{-40}$ and $|f^{ee}_3|\leq 1.4 \times 10^{-17}$ at the scale of $40$ mm (30 $\mu$eV).

\subsection{Geoelectrons}
The methods described in the two previous sections rely on experimental setups, where both the ``source'' of the exotic force and the ``detector'' sensitive to this force are situated in a laboratory. One may use another approach, where the ``source'' is located outside the laboratory. As an example of realisation of this idea we discuss the use of geoelectrons, i.e. polarized electrons within the Earth. 

The authors of Refs.\ \cite{Hun13, Hun14} constrained several spin-dependent and velocity-dependent potentials at planetary scales by comparing results of local Lorentz-invariance searches \cite{Hou03, Bro10} with an electron spin density map constructed by the authors. With the use of recent advances in fields such as geophysics, seismology, or mineral physics it was possible to model temperature, magnetic field, and density of unpaired electrons within the Earth, and to ultimately obtain a complete map of electron spin density. Then appropriate integrations over the whole planet volume yielded estimates for the possible influences of exotic interactions coming from geoelectrons. Finally, by comparing these estimates and the experimental data it was possible to obtain stringent constraints on various coupling constants \cite{Hun13, Hun14}, such as $|f^{ee}_2|\leq 5.7 \times 10^{-47}$ at the scale of 10 000 km ($1.2\times 10^{-12}$ eV).

\section{Summary and Outlook}\label{sec:sum}
In this brief paper, we provided a glimpse of the theory underlying the ongoing searches for exotic interactions and gave several examples of searches spanning a broad range of spatial scales, from the atomic subnanometer scale all the way to the planetary scale of tens of thousand kilometers. Such experiments provide a powerful way to look for physics beyond the standard model. At the same time, they constitute an indirect search for possible components of dark matter and dark energy. There are all indications that we will see significant improvement in the sensitivity of these methods in the coming years via a combination of improvements in the sensitivity of the experiments combined (where necessary) with higher-accuracy theory.

\section*{Acknowledgements}
We would like to thank our companions in a journey through the vast land of exotic interactions: Pavel Fadeev, Victor V. Flambaum, Mikhail G. Kozlov, Nathan Leefer, Szymon Pustelny, and Yevgeny V. Stadnik. Special thanks go to Derek F. Jackson Kimball for not only assisting us in the aforementioned expedition, but also for his insightful comments on this paper.

F.F. has been supported by the Polish Ministry of Science and Higher Education within the Diamond Grant (Grant No. 0143/DIA/2016/45). D.B. acknowledges the support of the European Research Council under the European Union's Horizon 2020 Research and Innovative Program under Grant agreement No. 695405 and by the DFG under the Reinhart Koselleck program.

\end{document}